\documentclass[aps,prl,reprint,superscriptaddress]{revtex4-1}

\usepackage{amsmath}
\usepackage{graphicx}

\begin{document}

\title{Exciton-Exciton Annihilation Is Coherently Suppressed in H-Aggregates,\\but Not in J-Aggregates}

\author{Roel Tempelaar}
\email{r.tempelaar@gmail.com}
\affiliation{University of Groningen, Zernike Institute for Advanced Materials, Nijenborgh 4, 9747 AG Groningen, The Netherlands}
\affiliation{Department of Chemistry, Columbia University, 3000 Broadway, New York, New York 10027, USA}

\author{Thomas L.~C.~Jansen}
\affiliation{University of Groningen, Zernike Institute for Advanced Materials, Nijenborgh 4, 9747 AG Groningen, The Netherlands}

\author{Jasper Knoester}
\email{j.knoester@rug.nl}
\affiliation{University of Groningen, Zernike Institute for Advanced Materials, Nijenborgh 4, 9747 AG Groningen, The Netherlands}

\begin{abstract}
We theoretically demonstrate a strong dependence of the annihilation rate between (singlet) excitons on the \textit{sign} of dipole-dipole couplings between molecules. For molecular H-aggregates, where this sign is positive, the phase relation of the delocalized two-exciton wavefunctions causes a destructive interference in the annihilation probability. For J-aggregates, where this sign is negative, the interference is constructive instead, as a result of which no such coherent suppression of the annihilation rate occurs. As a consequence, room temperature annihilation rates of typical H- and J-aggregates differ by a factor of $\sim$3, while an order of magnitude difference is found for low-temperature aggregates with a low degree of disorder. These findings, which explain experimental observations, reveal a fundamental principle underlying exciton-exciton annihilation, with major implications for technological devices and experimental studies involving high excitation densities. 
\end{abstract}

\maketitle

The annihilation between (singlet) excitons is a dominant contributor to the optoelectronic properties of materials at high excitation densities. It is considered a major loss mechanism in lasers based on organic thin films \cite{Baldo_02a} and polariton microcavities \cite{Akselrod_10a}, as well as organic light-emitting diodes \cite{Baldo_98a}. It is also an important factor impacting the excited state dynamics of single-walled carbon nanotubes \cite{Ying-Zhong_05a, Nguyen_11a, Moritsubo_10a} and inorganic monolayers \cite{Sun_14a}. At the same time, it has a functional purpose in the formation of interchain species \cite{Nguyen_00a} and separated charges \cite{Fuckel_09a, Gelinas_13a} in organic electronics. Exciton-exciton annihilation occurring in nonlinear spectroscopy at high fluences can complicate the interpretation of the measurements \cite{Valkunas_09a, Bruggemann_09a, Muller_10a}, while it also serves as a means to study the structure and functioning of materials \cite{DenHollander_83a, Fennel_15a}. In particular, it continues to find widespread application to determine exciton diffusion lengths through its imprints on laser fluence-dependent time-resolved spectroscopic measurements \cite{Barzda_01a, Shaw_10a, Cook_10a, Marciniak_11a, Tamai_14a, Lin_14a}.

Exciton-exciton annihilation (EEA) is commonly regarded as an incoherent, stochastic process, being described by the bi-molecular rate equation
\begin{align}
\Gamma=\alpha n^2,
\label{Eq_1}
\end{align}
with $\Gamma$ as the annihilation rate, $n$ as the exciton density, and $\alpha$ as a proportionality constant. A few theoretical studies \cite{Malyshev_99a, Ryzhov_01a, Bruggemann_01, May_14a, Hader_16a} have considered EEA beyond such a macroscopic description, and investigated the role of microscopic properties such as exciton coherence length \cite{Malyshev_99a, Ryzhov_01a} and relaxation pathways \cite{Bruggemann_01}. Nevertheless, our microscopic understanding of EEA remains limited, which hampers the rational design of materials with desirable EEA qualities. In particular, experiments have shown EEA to be much more effective in J-aggregates than in H-aggregates \cite{Khairutdinov_97a, Scheblykin_00a, King_07a, Ito_09a, Volker_14a}, for which a convincing explanation remains to be found.

Here, by applying a microscopic model, we demonstrate a dramatic dependence of EEA on the \textit{sign} of dipole-dipole couplings between molecules, $J$, which drives exciton delocalization. For H-aggregates, where $J>0$, the phase relation of the optically dominant, delocalized two-exciton wavefunctions contributes destructively to $\Gamma$. In contrast, no such destructive interference occurs for J-aggregates, for which $J<0$.

\begin{figure}
\centering
\includegraphics[scale=.714]{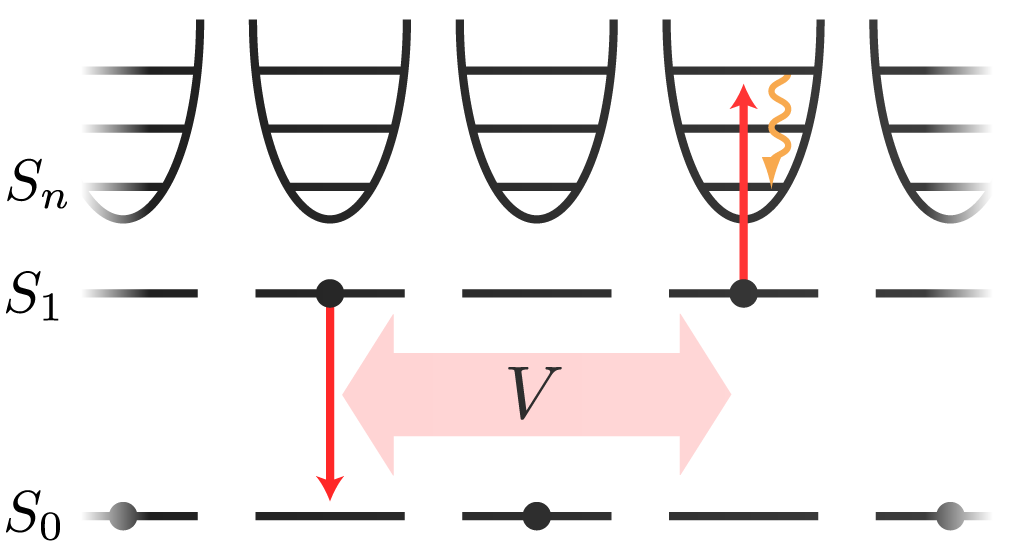}
\caption{Microscopic representation of exciton-exciton annihilation. Coupling ($V$) between nearby molecules in the $S_1$ state lowers one molecule to the $S_0$ state while promoting the other to $S_n$ (red arrows). Subsequently, phonon-assisted relaxation (yellow wiggling arrow) prohibits the regeneration of two $S_1$ excitations. Ultimately, the $S_n$ excitation decays back to $S_1$. Hence, the overall process corresponds to the loss of one $S_1$ excitation.}
\label{Fig1}
\end{figure}

Figure \ref{Fig1} provides a microscopic representation of EEA. Excitation energy is transferred resonantly between two nearby molecules in their $S_1$ excited state, lowering one molecule to the ground state ($S_0$) while promoting the other to a higher-lying singlet state ($S_n$), upon which phonon-assisted relaxation of $S_n$ occurs. If the associated relaxation rate ($\gamma$) is large compared to the resonant coupling between molecules, the regeneration of two $S_1$ states is prohibited, and the overall process corresponds to the effective loss of one excitation quantum. Furthermore, EEA can then be microscopically described by Fermi's Golden Rule, where the density of final states is replaced by $1/\gamma$ \cite{Ryzhov_01a},
\begin{align}
\Gamma=\frac{2\pi}{\hbar\gamma}\sum_{\mu,\nu}P_{\mu,\nu}\sum_m\vert\langle S_{n(m)}\vert H_\mathrm{a}\vert\Psi_{\mu,\nu}\rangle\vert^2.
\label{Eq_2}
\end{align}
Here, $S_{n(m)}$ represents a higher-lying singlet excitation at a molecule labeled $m$. Localization of this excitation can be neglected owing to the large relaxation rate, $\gamma$. $\Psi_{\mu,\nu}$ represents the eigenstates of the manifold of two $S_1$ excitations. The summation over the associated quantum numbers, $\mu$ and $\nu$, is weighed by the Boltzmann factor $P_{\mu,\nu}=e^{-\omega_{\mu,\nu}/k_\mathrm{B}T}/\sum_{\mu',\nu'}e^{-\omega_{\mu',\nu'}/k_\mathrm{B}T}$, with $\omega_{\mu,\nu}$ as the eigenenergy associated with $\Psi_{\mu,\nu}$. In second quantization, the annihilation Hamiltonian appearing in Eq.~\ref{Eq_2} is given by
\begin{align}
H_\mathrm{a}=\sum_{m_1,m_2}V_{m_1,m_2}b^\dagger_{n(m_1)}b_{1(m_1)}b_{1(m_2)}+\text{H.c.},
\label{Eq_4}
\end{align}
where $b_{1(m)}$ and $b_{n(m)}$ represent the Pauli annihilation operators for $S_1$ and $S_n$ excitations at molecule $m$, respectively, and $V_{m_1,m_2}$ represents the resonant coupling between the $S_1-S_n$ and $S_0-S_1$ transitions at molecules $m_1$ and $m_2$. The double summation in Eq.~\ref{Eq_4} implicitly excludes $m_1=m_2$, as will be the case for all other double summations appearing in this text.

\begin{figure}
\centering
\includegraphics{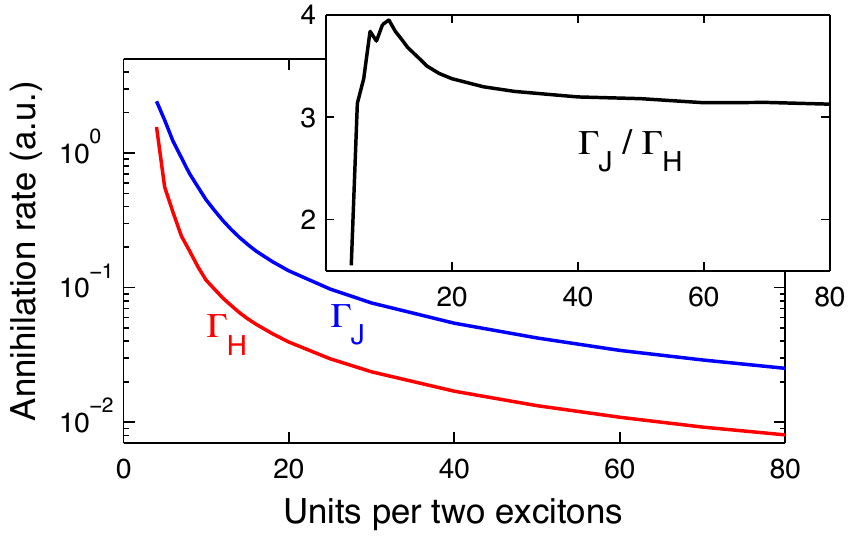}
\caption{Annihilation rates calculated using parameters typical for (linear) J- and H-aggregates, as a function of the number of molecular units per two excitons. Inset shows the ratio between the rates. (Irregular behavior observed using less than 10 units is owing to boundary effects.)}
\label{Fig2}
\end{figure}

Shown in Fig.~\ref{Fig2} are the calculated EEA rates for typical parameters representing linear J- and H-aggregates as a function of the number of molecular units in the aggregate. Imposing periodic boundaries, such molecular chains effectively represent extended aggregates with two $S_1$ excitons per molecule count. The two-exciton eigenstates and -energies, $\Psi_{\mu,\nu}$ and $\omega_{\mu,\nu}$, are obtained by solving the Schr\"odinger equation using the Hamiltonian
\begin{align}
H=\sum_m\epsilon_mb^\dagger_{1(m)}b_{1(m)}+\sum_{m_1,m_2}J_{m_1,m_2}b^\dagger_{1(m_1)}b_{1(m_2)},
\end{align}
where the first term contains the $S_0-S_1$ transition energies. Disorder in these energies is accounted for by drawing each $\epsilon_m$ randomly and independently from a normal distribution (centered at some $m$-independent value) with a standard deviation $\sigma=500$~cm$^{-1}$, while sampling over 20\;000 configurations. The second term accounts for dipole-dipole coupling between the $S_0-S_1$ transitions at molecules $m_1$ and $m_2$. Adopting the point-dipole approximation, and assuming all dipoles to be parallel, the coupling strength is given by $J_{m_1,m_2}=J_\mathrm{NN}/\vert m_1-m_2\vert^3$, using $J_\mathrm{NN}=\pm1000$~cm$^{-1}$. The couplings appearing in $H_\text{a}$ can likewise be regarded to be of dipolar form, and as such will differ from $J_{m_1,m_2}$ mostly by a constant prefactor. Since this difference will factor out in the equations under consideration, we simply set $V_{m_1,m_2}=J_{m_1,m_2}$. Lastly, the thermal distribution $P_{\mu,\nu}$ is taken for a temperature of $T=300$ K.

Figure \ref{Fig2} demonstrates an expected monotonous decrease of the EEA rate with increasing aggregate length, or decreasing excitation density. However, throughout, the rate for H-aggregates is found to be consistently and significantly lower than the equivalent for J-aggregates. This is particularly evident when considering the rate ratio, $\Gamma_\text{J}/\Gamma_\text{H}$, which rapidly converges to a value of $\sim$3.1. This pronounced difference is obtained by simply inverting the sign of the dipole-dipole couplings from $J_\mathrm{NN}=+1000$~cm$^{-1}$ (H-aggregates) to $J_\mathrm{NN}=-1000$~cm$^{-1}$ (J-aggregates), and suggests a fundamental principle affecting EEA that goes beyond a macroscopic representation of this process.

In order to understand the continuous difference between $\Gamma_\text{J}$ and $\Gamma_\text{H}$, it is instructive to consider the limiting case of zero temperature ($T=0$~K) and without disorder ($\sigma=0$). This case can be solved analytically, yielding $\Gamma_\text{J}=4\Gamma_\text{H}$ (assuming nearest-neighbor interactions for $J_{m_1,m_2}$, but point-dipole interactions for $V_{m_1,m_2}$, see Supplemental Material). Importantly, in this case only the lowest-energy (band-bottom) eigenstate contributes to the EEA rate. The J- and H-aggregate band-bottom eigenstates are expanded in the local basis as $\vert\Psi_\text{J/H}\rangle=\sum_{m_1>m_2}c_{m_1,m_2}^\text{J/H}\vert m_1,m_2\rangle$, where $\vert m_1,m_2\rangle$ represents a pair of $S_1$ excitations at molecules $m_1$ and $m_2$. It is helpful to define symmetrized wavefunction coefficients as $d_{m_1,m_2}^\text{J/H}\equiv\Theta(m_1-m_2)c_{m_1,m_2}^\text{J/H}+\Theta(m_2-m_1)c_{m_2,m_1}^\text{J/H}$, where $\Theta(m)$ is the Heaviside step function (so that $d_{m_1,m_2}^\text{J/H}=d_{m_2,m_1}^\text{J/H}$). It can be shown (see Supplemental Material) that under these conditions the EEA rate is given by
\begin{align}
\Gamma_\text{J/H}=\frac{2\pi}{\hbar\gamma}\sum_{m_1}\vert\sum_{m_2}V_{m_1,m_2}d_{m_1,m_2}^\text{J/H}\vert^2.
\label{Eq_3}
\end{align}
Hence, the EEA rate scales as the square of the coherent sum over $m_2$ of the product $V_{m_1,m_2}d_{m_1,m_2}^\text{J/H}$. Note that this sum is independent of $m_1$, owing to the periodic boundaries imposed and the absence of disorder. Furthermore, the coupling $V_{m_1,m_2}$ is mono-signate and scales as $1/\vert m_1-m_2\vert^3$, leaving the coefficients to determine the difference between J- and H-aggregates.

\begin{figure}
\centering
\includegraphics{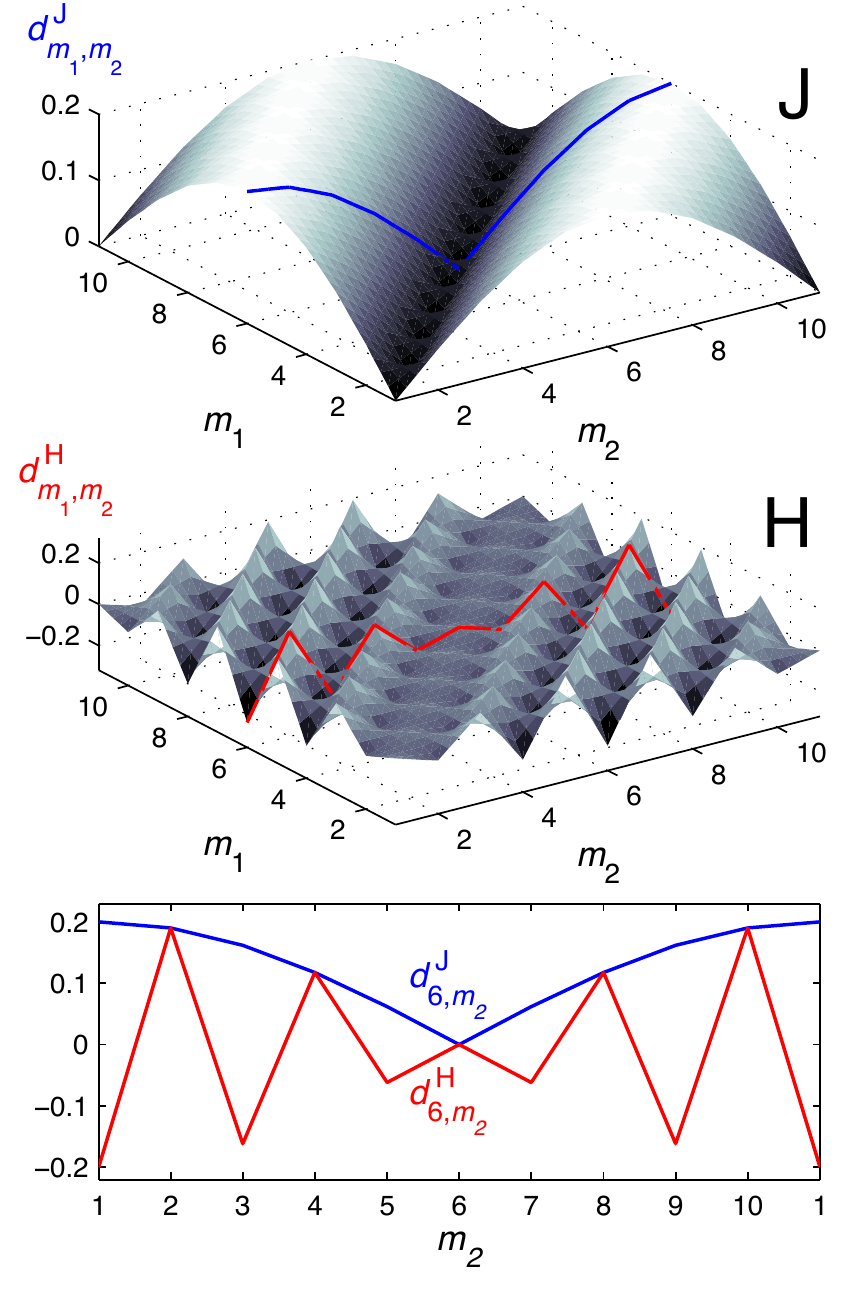}
\caption{Symmetrized wavefunction coefficients of the band-bottom eigenstates of the two-exciton Hamiltonian for J- (top) and H-aggregates (middle) consisting of 10 molecules, together with a slice taken at $m_1=6$ (bottom).}
\label{Fig3}
\end{figure}

The symmetrized coefficients $d_{m_1,m_2}^\text{J/H}$ are plotted in Fig.~\ref{Fig3} for an aggregate of length 10. From this figure, the fundamental difference between J- and H-aggregates becomes evident. For J-aggregates, the coefficients are in-phase for all values of $m_2$. As a result, they constructively contribute to the coherent sum in Eq.~\ref{Eq_3}. For H-aggregates, on the other hand, the coefficients are sign-alternating with $m_2$. This phase relation, combined with the long range of $V_{m_1,m_2}$, results in a destructive interference in Eq.~\ref{Eq_3}. This behavior is akin (that is, formally similar) to super-radiance and sub-radiance observed upon intra-band relaxation in J- and H-aggregates, respectively \cite{Spano_10a}. Similarly to this phenomenon, the responsible destructive interference is maximal only for the band-bottom state in the absence of disorder, but the effect is nevertheless retained when disorder is present and at finite temperatures.

\begin{figure}[t]
\centering
\includegraphics{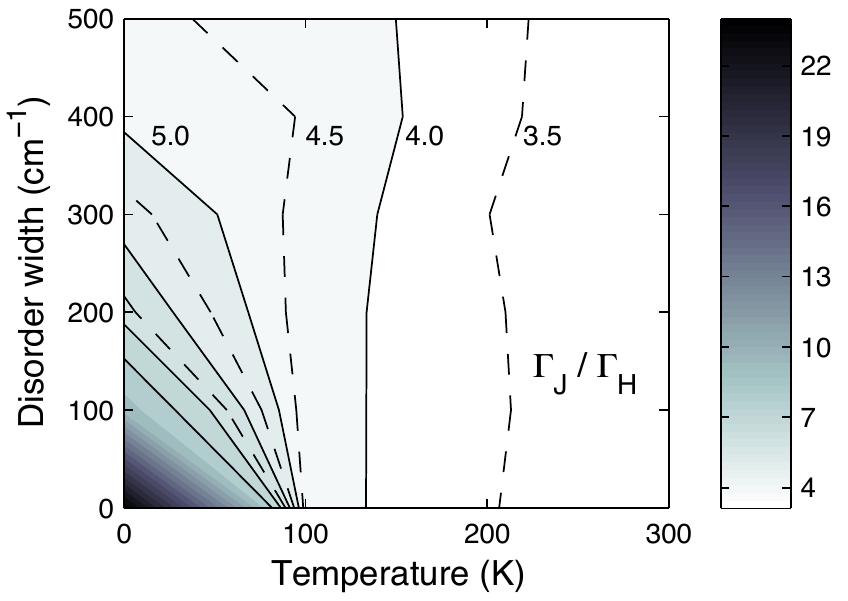}
\caption{Ratio of the annihilation rates of J- and H-aggregates as a function of temperature and disorder width, calculated using 80 molecules per two excitons.}
\label{Fig4}
\end{figure}

Shown in Fig.~\ref{Fig4} is the ratio of $\Gamma_\text{J}$ and $\Gamma_\text{H}$ as a function of the disorder width ($\sigma$) and temperature ($T$), calculated for linear aggregates consisting of 80 molecules. Results for $\sigma=100$~cm$^{-1}$, $200$~cm$^{-1}$, $300$~cm$^{-1}$, $400$~cm$^{-1}$, and $500$~cm$^{-1}$ are averaged over 4000, 8000, 12\;000, 16\;000, and 20\;000 configurations, respectively. This figure demonstrates that the contrasting behavior of J- and H-aggregates is not unique to disorder-free systems at low temperature, but applies equally well for disordered aggregates over all physically relevant temperatures. Note that with $J_{m_1,m_2}$ taken in the point-dipole approximation, the annihilation ratio at low values of $T$ and $\sigma$ diverges with increasing aggregate length. This is in contrast to the disorder-free case at zero temperature with $J_{m_1,m_2}$ limited to nearest-neighbors, for which the ratio asymptotically approaches 4, as shown in the Supplemental Material. Although the physical origin of this difference is beyond the scope of the current work, we have performed additional calculations (not shown here) demonstrating that an extension of $J_{m_1,m_2}$ beyond nearest neighbors yields a further suppression of annihilation for H-aggregates while yielding an enhancement for J-aggregates, which accounts for this observation.

The above demonstration of the coherent suppression of EEA in H-aggregates adds to a recent trend connecting macroscopic material properties to the phase of quantum excitations, and its sensitivity to the sign of intermolecular couplings, through microscopic modeling. For example, recent studies have found that the interference between dipole-dipole couplings and short-ranged charge transfer interactions underlies the diversity in absorption spectra displayed by chemically near-identical molecular crystals \cite{Yamagata_14a}, and offers the possibility to control the exciton mobility in such materials \cite{Hestand_15a}. Other studies demonstrated the importance of wavefunction delocalization to charge recombination at molecular heterojunctions \cite{Bittner_14a}, and the crucial role of the signs of charge-transfer integrals in the suppression of this loss mechanism \cite{Tempelaar_16a, Tempelaar_16b}. Lastly, interference between charge transfer interactions is shown to be of key importance in singlet exciton fission.\cite{Petelenz_16a}

The implications of our findings to technological applications and experiments involving high excitation densities are straightforward. For molecular devices where EEA is undesirable, the selective use of H-type materials (i.e., having predominantly negative dipole-dipole couplings) is a possible means to minimize this loss mechanism. For devices where EEA serves a functional purpose, on the other hand, J-type materials are to be preferred. Our theory provides a plausible interpretation of the aforementioned experiments observing a higher EEA rate in J-aggregates compared to H-aggregates \cite{Khairutdinov_97a, Scheblykin_00a, Ito_09a, Volker_14a}. Generally, it predicts the contribution of EEA to nonlinear spectroscopy to be significantly smaller for H-type materials than for J-type materials. The latter has an important consequence for studies seeking to determine exciton diffusion lengths using fluence-dependent time-resolved spectroscopy, since this approach likely yields significant underestimates for H-aggregates. As such, it is of great interest to assess the accuracy of such studies through a comparison with more direct methods of determining diffusion lengths, such as optical absorption microscopy \cite{Devadas_15a, Wan_17a}. Drawing such a comparison will simultaneously offer a firm experimental verification of the theory proposed in this work.

In summary, we have demonstrated that the sign of dipole-dipole couplings between molecules has a profound impact on the annihilation rate between (singlet) excitons through interference of the phase relations of the two-exciton wavefunctions. In H-aggregates, with positive couplings, this interference is destructive as a result of which the rate is significantly suppressed. For J-aggregates, where couplings are negative, no such coherent suppression occurs. This gives rise to a factor of $\sim$3 difference between annihilation rates for typical J- and H-aggregates. These findings explain experimental observations, and open an avenue for the rational design of materials with desirable annihilation qualities.

R.T. acknowledges The Netherlands Organisation for Scientific Research NWO for support through a Rubicon grant. 

\bibliographystyle{../apsrev4-1Roel}

%

\end{document}